\documentclass[preprint]{aastex}
\def\simgr{\,\hbox{\hbox{$ > $}\kern -0.8em \lower 1.0ex\hbox{$\sim$}}\,}
\def\simle{\,\hbox{\hbox{$ < $}\kern -0.8em \lower 1.0ex\hbox{$\sim$}}\,}

\shortauthors{KAPUSTA}
\shorttitle{RXS J053234.9+624755}
\begin{document}
\title{ Orbital Period of the Dwarf Nova RXS J053234.9+624755
\footnote{Based on observations obtained at the MDM Observatory, operated by
Dartmouth College, Columbia University, Ohio State University, and
the University of Michigan.}
}

\author{Ann B. Kapusta and John R. Thorstensen}
\affil{Department of Physics and Astronomy\\
6127 Wilder Laboratory, Dartmouth College\\
Hanover, NH 03755-3528;\\
ann.kapusta@dartmouth.edu}

\begin{abstract}

We report spectroscopy of the newly discovered SU Ursae Majoris dwarf nova
identified with the x-ray source  RXS J053234.9+624755.  Radial velocities of
the H$\alpha$ emission line in the quiescent state give an orbital period of
0.05620(4) d (80.93 min),  which is among the shortest for SU UMa stars with
determined periods.  We also report UBVI magnitudes of the quiescent dwarf nova
and surrounding stars.  Using a previous measurement of the superhump period,
we find the fractional superhump excess $\epsilon$ to be 0.016(4), which is not
atypical of dwarf novae in this period range.  
\end{abstract}


\section{Introduction}

Cataclysmic Variables (CVs) are close binary systems that consist of an
accreting white dwarf (the primary) and a secondary component that usually
resembles a main-sequence star.  \citet{warn} comprehensively reviews CVs.
Dwarf novae, or U Geminorum stars, are a subclass of CVs that can be further 
subclassified based on their outburst behavior.  The SU Ursae Majoris stars 
(referred to as UGSU)
form a subclass of dwarf novae that undergo occasional superoutbursts in
addition to normal outbursts.  Superoutbursts occur less frequently than normal
outbursts, but are brighter and last longer.  During a superoutburst,
characteristic oscillations called ``superhumps''  develop in the light curves.
The measured superhump period, $P_{\rm sh}$, is usually a few percent longer
than the measured orbital period, $P_{\rm orb}$.  Almost all known SU UMa stars
have  $P_{\rm orb} < 2$ hr.  The superhump period excess, $\epsilon =$[($P_{\rm
sh}-P_{\rm orb}$/$P_{\rm orb}$], is an important quantity for these stars.
\citet{patt01} demonstrated that $\epsilon$ correlates well with the mass ratio
$q=M_2/M_1$, which is otherwise difficult to obtain.

Another class of dwarf novae are the WZ Sagittae stars (UGWZ), which are
extreme examples of the SU UMa-type stars.  WZ Sge stars exhibit
large-amplitude outbursts ($\geq$ 6 mag) that occur less frequently than the
those of the UGSU (\citealt{osaki}).  WZ Sge stars do exhibit superhumps in
their light curves, but do not appear to have any `normal' outbursts.

Here we present observations of the dwarf nova identified with the ROSAT x-ray
source RXS J053234.9+624755 (hereafter RX0532+62).  The discovery of this star,
which lies in Camelopardalis, is described by \citet{poy}.  \citet{bernhard}
classified it as a U Gem-type dwarf novae with a recurrence time scale of 133
days.  The relatively frequent outbursts show that this star is not a UGWZ.
RX0532+62 underwent a well-observed superoutburst in 2005 March.  During this
outburst a superhump period,  $P_{\rm sh}$, of  0.0571(2) d (82.2 mins) was
reported by Tonny Vanmunster (CBA Belgium) \footnote{``Detection of Superhumps
in the CV 1RXS J053234.9+624755'' can be found at
http:$\rm//users.skynet.be/fa079980/cv\_2005/1RXSJ053234\_2005\_mar\_18.htm$};
\citet{poy} independently found a similar, but less accurate, value of $P_{\rm
sh}$.  We undertook observations of this star to independently determine the
orbital period, $P_{\rm orb}$, and allow determination of the superhump period
excess, $\epsilon$.

\section{Observations}

We took spectra using the MDM Observatory 1.3m McGraw-Hill telescope on Kitt
Peak during two observing runs in 2005 September and 2006 January. We used the
Mark III spectrograph and a thinned $1024^2$ SITe CCD detector, with a 600 line
mm$^{-1}$ grism that gave spectra spanning from 4500 \AA\ to 7000 
\AA\ at 2.3 \AA\ pixel$^{-1}$.  In all, 106 spectra of RX0532+62 were
taken with typical exposure times of 480 s, along with frequent comparison
spectra to maintain the wavelength calibration.  At the beginning and end of
every photometric night, we observed flux standards to determine an absolute
flux calibration.  Table 1 gives a journal of observations.  The system was
in quiescence for all the observations reported here.

We also obtained direct images with the MDM 2.4m Hiltner Telescope.  A SITe
$2048^2$ CCD with an image scale of 0.275 arcsec pixel$^{-1}$ was used, with
filters matching the Johnson UBV and Kron-Cousins I passbands.  The stars in
the field were calibrated using \citet{landolt92} standards.  Exposures were
obtained while the sky appeared photometric, and the reduced images were
processed using the DAOphot program as implemented in IRAF\footnote{The Image
Reduction and Analysis Facility software is written and maintained by the IRAF
programming group located at the National Optical Astronomy Observatory (NOAO).
Software and information can be found at http://iraf.noao.edu.}
(\citealt{stetson}).  Positions for stars in the field were derived by fitting
a plate model to 39 USNO A2.0 stars (\citealt{USNOA2.0}); the fit had an RMS
error of 0.37 arcsec.  Table 2 shows the celestial positions, magnitudes, and
colors of RX0532+62 and the field stars.  Figure 1 shows the field around
RX0532+62 along with the measured V magnitudes. 

We reduced the spectra using standard IRAF routines, except for the extraction
of one-dimensional spectra from the two-dimensional images.  For this, we used
an original implementation of the optimal extraction algorithm developed by
\citet{horne}; the primary advantage over the IRAF ${\it apsum}$ routine was an
improved rejection of bad pixels.  The time averaged and flux-calibrated
spectrum of RX0532+62 is shown in Figure 2.  The spectrum appears typical of
dwarf novae, showing strong broad emission lines.  The double peaks in the
emission lines imply that the orbital inclination is not too far from edge-on.
To measure the emission line radial velocities we used the convolution method
described by \citet{sy80}.  The steep sides of the line profile were measured
by convolving the line with a function consisting of positive and negative
gaussians displaced by an adjustable separation.  The emission lines' widths
and strengths were measured in the time-average spectra; the results are given
in Table 3.  

To search for $P_{\rm orb}$ in the radial velocities, we used the
``residualgram'' method as described by \citet{thor96}.  Figure 3 shows the
result for the 2005 September data.  Table 4 gives the parameters of the best
sine fits of the form $v(t) = \gamma + K\sin[2\pi(t-T_{o})/P]$, and the rms
scatter $\sigma$ around the best fits.  Figure 4 shows the velocities folded on
the period adopted from the combined (2005 September, 2006 January) data,
together with the best-fitting sinusoid. The 2005 September data did not
unambiguously determine the correct choice of daily cycle count because of the
limited hour angle coverage available early in the observing season.  The 2006
January data were taken in order to resolve the ambiguity, but for unknown
reasons the velocities had greater scatter than the 2005 September data.  We
also measured velocities of the H$\beta$ emission in an attempt to resolve the
daily cycle count.  The H$\beta$ velocities corroborated the H$\alpha$
measurements, but the period remained ambiguous.  However, we know $P_{\rm
sh}$, and that the $P_{\rm orb}$ of an SU UMa-type should be a few percent less
than $P_{\rm sh}$.  This guides our choice of cycle count, which yields
0.05620(4) d for the 2005 September data.  The run-to-run cycle count is
ambiguous, but periods consistent with all the data are given by $$P_{\rm
orb}={130.588 \pm 0.015 \over 2324 \pm 6}{\rm \ \ d},$$ 
where the numerator is the measured
interval between blue-to-red crossings of the H$\alpha$ emission velocities 
determined from our two observing runs, and the denominator is constrained 
to integer values.

\section{Discussion}

Combining our $P_{\rm orb}$ with the previously measured $P_{\rm sh}$, we find
$\epsilon = 0.016(4)$.  \citet{patt03} plot log($\epsilon$) against log($P_{\rm
orb}$) for a large number of systems with hydrogen-rich secondaries.  On this
plot, RX0532+62 lies near the short-period end, where systems appear to be
evolving through the period minimum.  These occupy a relatively wide range of
$\epsilon$ values.  RX0532+62 is toward the top of this range, as if it has
evolved into the turnaround region relatively recently.  \citet{patt01} fits
the empirical relation between $\epsilon$ and the mass ratio $q$ as
$\epsilon(q)=0.216q$.  Using our $\epsilon$ value, we determine $q=0.074(19)$,
which is typical for dwarf novae with similar periods\footnote{The uncertainty in
$q$ is computed here using only the uncertainty in $\epsilon$; imperfections
in the empirical relation are ignored.}. We conclude that
RX0532+62 is a typical SU UMa star lying near the minimum period for
hydrogen-rich secondary systems.

{\it Acknowledgments}: We thank Holly Sheets for taking the 2006 January 
spectra.  The  National Science Foundation funded this research
through award AST-0307413 and an REU supplement to that award.  Travel for Ann
Kapusta was made possible by a generous gift from Claudia and Jay Weed.

\clearpage

\begin{deluxetable}{lccc}
\tablewidth{0pt}
\tablecolumns{4}
\tablecaption{Journal of Observations}
\tablehead{
\colhead{Dates(UT)} &
\colhead{$N$\tablenotemark{a}} &
\colhead{HA Start\tablenotemark{b}} &
\colhead{HA End\tablenotemark{c}} 
}
\startdata
\cutinhead{Spectroscopy (1.3m)}
2005 Sep 13 & 23 & $ -5:01$ & $ -1:32$ \\ 
2005 Sep 14 & 17 & $ -4:10$ & $ -1:41$ \\ 
2005 Sep 15 & 13 & $ -3:01$ & $ -1:15$ \\ 
2005 Sep 16 & 18 & $ -5:07$ & $ -2:38$ \\ 
2006 Jan 22 & 10 & $ +2:16$ & $ +4:10$ \\ 
2006 Jan 23 & 25 & $ -2:55$ & $ +0:33$ \\ 
\cutinhead{UBVI Direct Imaging (2.4m)}
2005 Sep 14 & 4 & $-$1:03  & $-$0:59 \\
\enddata
\tablenotetext{a} {Number of Exposures}
\tablenotetext{b} {Hour angle at the midpoint of the first exposure in hours and minutes}
\tablenotetext{c} {Hour angle at the midpoint of the exposure in hours and minutes}
\end{deluxetable}

\clearpage

\begin{deluxetable}{llrrrr}
\tablewidth{0pt}
\tablecolumns{6}
\tablecaption{Filter Photometry}
\tabletypesize{\small}
\tablehead{
\colhead{$\alpha$\tablenotemark{a}} &
\colhead{$\delta$\tablenotemark{a}} &
\colhead{$U-B$} &
\colhead{$B-V$} &
\colhead{$V$} &
\colhead{$V-I$} \\
}

\startdata
\cutinhead{Field stars}
 5:32:15.52 & +62:46:28.3 & $  1.24 \pm  0.02 $& $  1.19 \pm  0.01 $& $ 13.87 \pm  0.00 $& $  1.21 \pm  0.01 $ \\ 
 5:32:15.99 & +62:47:23.7 & $  0.24 \pm  0.03 $& $  0.73 \pm  0.01 $& $ 15.94 \pm  0.01 $& $  0.84 \pm  0.01 $ \\ 
 5:32:17.30 & +62:47:17.9 & $  0.62 \pm  0.14 $& $  0.91 \pm  0.02 $& $ 16.98 \pm  0.01 $& $  1.04 \pm  0.01 $ \\ 
 5:32:23.84 & +62:49:30.5 & $  0.38 \pm  0.11 $& $  0.88 \pm  0.03 $& $ 17.00 \pm  0.02 $& $  0.94 \pm  0.03 $ \\ 
 5:32:26.39 & +62:48:49.9 & $  0.84 \pm  0.06 $& $  1.09 \pm  0.01 $& $ 15.59 \pm  0.01 $& $  1.18 \pm  0.01 $ \\ 
 5:32:29.45 & +62:46:02.9 & $  0.21 \pm  0.02 $& $  0.68 \pm  0.01 $& $ 14.97 \pm  0.00 $& $  0.78 \pm  0.01 $ \\ 
 5:32:30.82 & +62:49:49.3 & $  0.45 \pm  0.01 $& $  0.85 \pm  0.01 $& $ 14.21 \pm  0.00 $& $  0.86 \pm  0.00 $ \\ 
 5:32:30.96 & +62:50:06.6 & $  0.17 \pm  0.26 $& $  0.92 \pm  0.06 $& $ 17.86 \pm  0.02 $& $  1.04 \pm  0.03 $ \\ 
 5:32:31.46 & +62:47:58.1 & $  0.20 \pm  0.03 $& $  0.75 \pm  0.01 $& $ 15.62 \pm  0.01 $& $  0.84 \pm  0.01 $ \\ 
 5:32:32.18 & +62:49:39.0 & $  1.42 \pm  0.04 $& $  1.29 \pm  0.01 $& $ 14.40 \pm  0.00 $& $  1.29 \pm  0.00 $ \\  
 5:32:37.91 & +62:49:54.1 & $  0.27 \pm  0.05 $& $  0.78 \pm  0.01 $& $ 16.12 \pm  0.01 $& $  0.85 \pm  0.01 $ \\ 
 5:32:39.68 & +62:46:29.3 & $  0.40 \pm  0.01 $& $  0.84 \pm  0.00 $& $ 14.31 \pm  0.00 $& $  0.94 \pm  0.01 $ \\ 
 5:32:39.69 & +62:50:10.7 & $  0.44 \pm  0.19 $& $  0.87 \pm  0.03 $& $ 17.45 \pm  0.01 $& $  0.98 \pm  0.02 $ \\ 
 5:32:40.04 & +62:47:45.1 & $  0.89 \pm  0.38 $& $  1.30 \pm  0.04 $& $ 17.32 \pm  0.01 $& $  1.34 \pm  0.02 $ \\ 
 5:32:40.46 & +62:46:23.5 & $  0.96 \pm  0.17 $& $  0.99 \pm  0.03 $& $ 16.91 \pm  0.01 $& $  1.06 \pm  0.01 $ \\ 
 5:32:44.65 & +62:45:39.9 & $  0.39 \pm  0.01 $& $  0.84 \pm  0.01 $& $ 14.13 \pm  0.01 $& $  0.87 \pm  0.01 $ \\ 
 5:32:46.49 & +62:45:48.1 & $  0.17 \pm  0.04 $& $  0.80 \pm  0.01 $& $ 15.84 \pm  0.01 $& $  0.84 \pm  0.01 $ \\ 
 5:32:47.28 & +62:47:55.8 & $  0.67 \pm  0.24 $& $  0.91 \pm  0.04 $& $ 17.45 \pm  0.01 $& $  0.97 \pm  0.02 $ \\ 
 5:32:48.62 & +62:47:20.6 & $  1.46 \pm  0.65 $& $  1.13 \pm  0.05 $& $ 17.74 \pm  0.02 $& $  1.21 \pm  0.02 $ \\ 
 5:32:49.35 & +62:47:35.3 & $  0.15 \pm  0.16 $& $  0.81 \pm  0.04 $& $ 17.64 \pm  0.01 $& $  0.95 \pm  0.02 $ \\ 
 5:32:49.44 & +62:48:22.9 & $  0.21 \pm  0.04 $& $  0.74 \pm  0.01 $& $ 16.05 \pm  0.01 $& $  0.80 \pm  0.01 $ \\ 
 5:32:49.82 & +62:48:12.4 & $  1.12 \pm  0.18 $& $  1.16 \pm  0.02 $& $ 16.45 \pm  0.01 $& $  1.22 \pm  0.01 $ \\ 
 5:32:50.98 & +62:46:16.5 & $  0.10 \pm  0.01 $& $  0.61 \pm  0.01 $& $ 13.47 \pm  0.01 $& $  0.63 \pm  0.01 $ \\ 
 5:32:52.02 & +62:46:18.8 & $  0.19 \pm  0.13 $& $  0.88 \pm  0.02 $& $ 16.70 \pm  0.01 $& $  0.94 \pm  0.01 $ \\ 
 5:32:52.04 & +62:49:10.8 & $  0.46 \pm  0.10 $& $  0.97 \pm  0.02 $& $ 16.78 \pm  0.01 $& $  1.02 \pm  0.02 $ \\ 
 5:32:54.02 & +62:48:58.2 & $  1.28 \pm  0.55 $& $  1.26 \pm  0.06 $& $ 17.78 \pm  0.02 $& $  1.42 \pm  0.02 $ \\
\cutinhead{Variable star}
 5:32:33.88 & +62:47:52.5 & $ -1.19 \pm  0.01 $& $  0.01 \pm  0.01 $& $ 16.25 \pm  0.01 $& $  0.65 \pm  0.01 $ \\

\enddata
\tablenotetext{a}{Coordinates referred to the ICRS(i.e. J2000), and are from a fit to
39 USNO A2.0 stars, with a scatter of 0.37 arcsec.  Right ascensions are measured in hours,
minutes, and seconds and the declinations are in degrees, minutes, and 
seconds.}
\end{deluxetable}

\clearpage

\begin{deluxetable}{lrcc}
\tablewidth{0pt}
\tablecolumns{4}
\tablecaption{Emission Features}
\tablehead{
\colhead{Feature} &
\colhead{E.W.\tablenotemark{a}} &
\colhead{Flux\tablenotemark{b}}  &
\colhead{FWHM \tablenotemark{c}} \\
 &
\colhead{(\AA )} &
\colhead{} &
\colhead{(\AA)} \\
}

\startdata
\cutinhead{RX 0532+62}
            H$\beta$ & $113$ & $1844$ & 32 \\ 
  HeI $\lambda 4921$ & $ 10$ & $162$ & 40 \\ 
  HeI $\lambda 5015$ & $ 12$ & $184$ & 37 \\ 
   Fe $\lambda 5169$ & $  9$ & $139$ & 36 \\ 
  HeI $\lambda 5876$ & $ 43$ & $524$ & 40 \\ 
           H$\alpha$ & $160$ & $2004$ & 35 \\ 
  HeI $\lambda 6678$ & $ 20$ & $236$ & 46 \\

\enddata
\tablenotetext{a}{Emission equivalent widths are counted as positive.}
\tablenotetext{b}{Absolute line fluxes in units of 10$^{-16}$ erg cm$^{-2}$ s$^{-1}$.  
These are uncertain by a factor of about
2, but relative fluxes of strong lines
are estimated accurate to $\sim 10$ per cent.}
\tablenotetext{c}{From Gaussian fits.}
\end{deluxetable}

\clearpage

\begin{deluxetable}{lllrrcc}
\tablecolumns{7}
\tabletypesize{\small}
\tablewidth{0pt}
\tablecaption{Fit to the Radial Velocities}
\tablehead{
\colhead{Data Set} &
\colhead{$T_0$\tablenotemark{a}} & 
\colhead{$P$} &
\colhead{$K$} & 
\colhead{$\gamma$} & 
\colhead{$N$} &
\colhead{$\sigma$\tablenotemark{b}}  \\ 
\colhead{} &
\colhead{} &
\colhead{(days)} & 
\colhead{(km s$^{-1}$)} &
\colhead{(km s$^{-1}$)} & 
\colhead{} &
\colhead{(km s$^{-1}$)} \\
}
\startdata
Combined & 53629.9522(15) & [0.0561950]\tablenotemark{c} &  38(6) & $-0(5)$ & 106 &  23 \\ 
2005 September & 53627.9854(10) & 0.05620(4) &  42(4) & $ 7(3)$ & 71 &  19 \\ 
2006 January & 53758.580(2) & 0.0557(3) &  38(10) & $-9(7)$ & 35 &  23 \\
\enddata
\tablecomments{Parameters of least-squares sinusoid fits to the radial
velocities, of the form $v(t) = \gamma + K \sin(2 \pi(t - T_0)/P$.}
\tablenotetext{a}{HJD - 2452000.}
\tablenotetext{b}{Root-mean-square residual of the fit.}
\tablenotetext{c}{This period is based on an arbitrary choice of cycle count 
between 2005 September and 2006 January.}
\end{deluxetable}

\clearpage

\begin{figure}
\caption{Finding chart with standard magnitudes.  The image is approximately 4.5 arc minutes square.  The labels are V-Magnitudes of stars in the field.
}
\end{figure}

\clearpage

\begin{figure}
\plotone{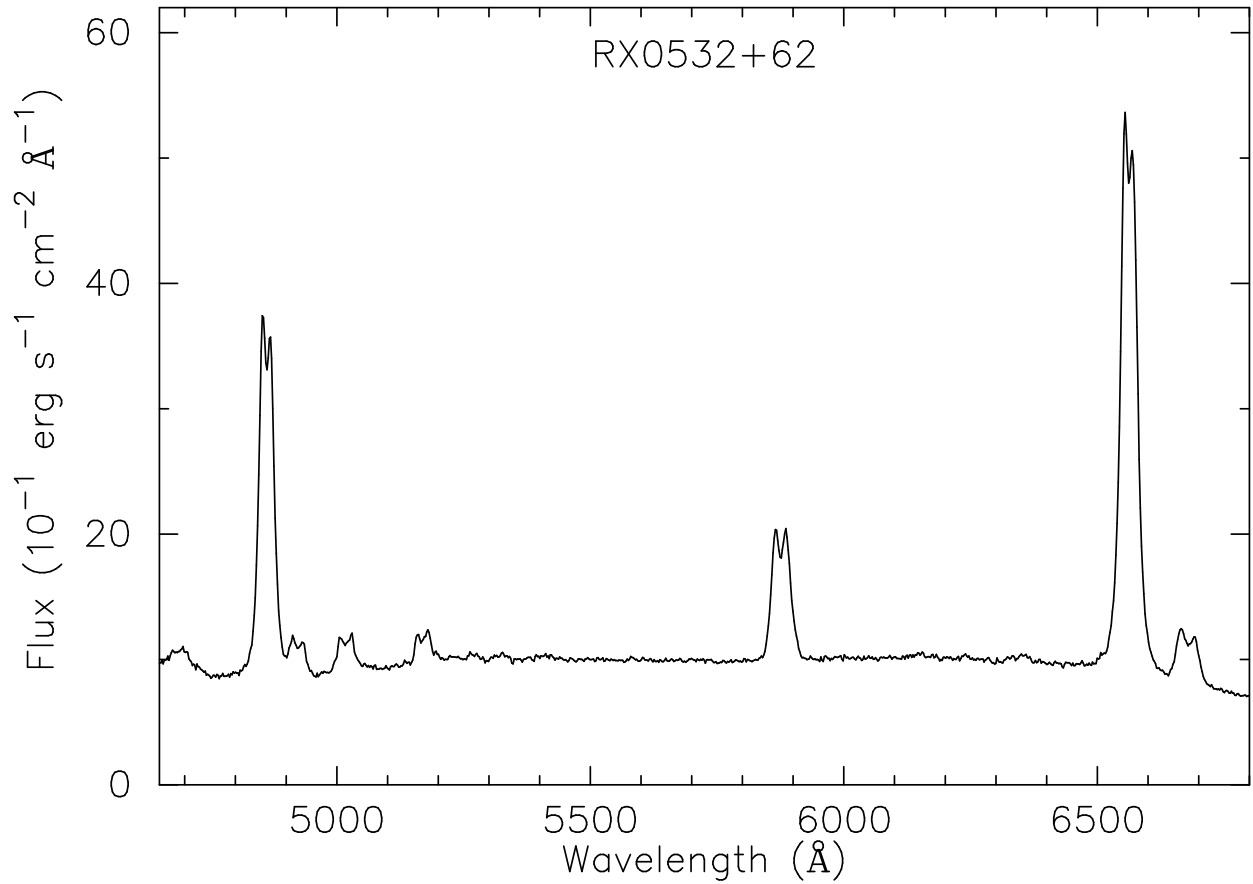}
\caption{Mean spectrum of RX0532+62 from 2005 September.  The 2006 
January average appeared nearly identical.
}
\end{figure}

\clearpage

\begin{figure}
\plotone{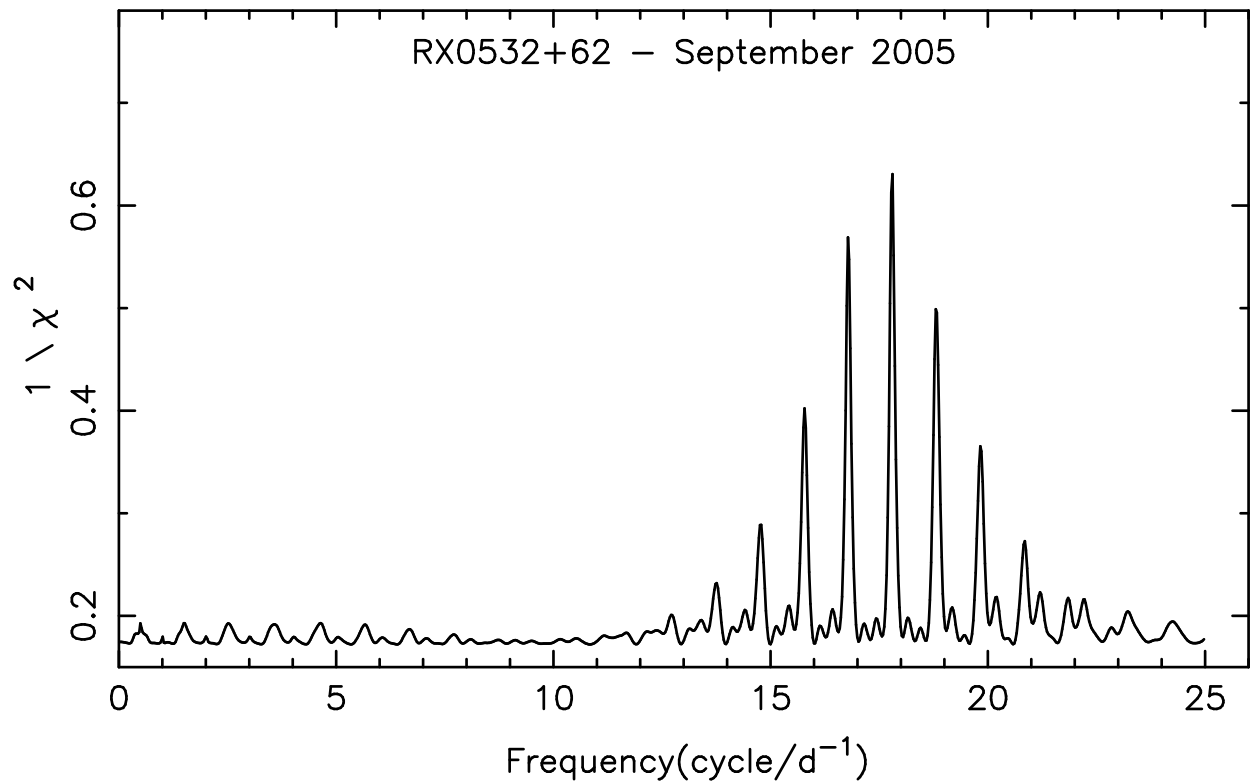}
\caption{Results of the period search on the H$\alpha$ radial velocities of RX0532+62.  The highest peak corresponds to the adopted $P_{\rm orb}$.
}
\end{figure}

\clearpage

\begin{figure}
\plotone{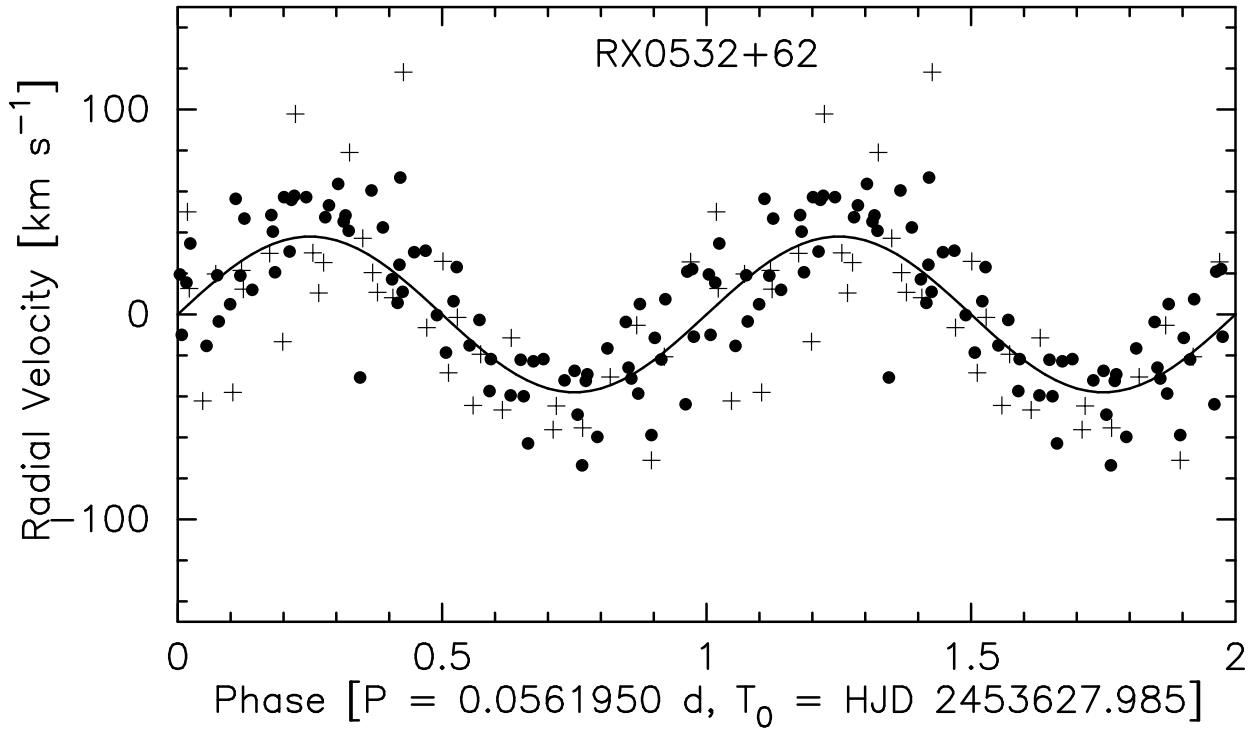}
\caption{Emission-line radial velocities as a function of orbital phase, together with the best-fitting
sinusoid.  The data and fit are repeated for a second cycle
to preserve continuity.  The pluses show velocities from 
2006 January, and the filled circles are from 2005
September.  The gross period is determined by the 2005 September data and comparison to the superhump period, but the precise period used here is based on an arbitrary choice of cycle count between the 2005 September and the 2006 January observing runs.
}
\end{figure}

\end{document}